\documentclass[prb,reprint,twocolumn,superscriptaddress,showpacs,amsmath,amssymb]{revtex4}
\usepackage{graphicx}% Include figure files
\usepackage{dcolumn}% Align table columns on decimal point
\usepackage{bm}% bold math
\usepackage{hyperref}
\usepackage{color}
\begin{document}

%\preprint{APS/123-QED}

\title{Electronic and magnetic structure of epitaxial NiO/Fe$_3$O$_4$(001) heterostructures grown on MgO(001) and Nb-doped SrTiO$_3$(001)}

\author{K.~Kuepper}\email{kkuepper@uos.de}
\affiliation{Department of Physics and Center of Physics and Chemistry of New Materials, Osnabr{\"u}ck University, 49076 Osnabr{\"u}ck, Germany}

\author{O.~Kuschel}
\affiliation{Department of Physics and Center of Physics and Chemistry of New Materials, Osnabr{\"u}ck University, 49076 Osnabr{\"u}ck, Germany}

\author{N.~Path{\'{e}}}
\affiliation{Department of Physics and Center of Physics and Chemistry of New Materials, Osnabr{\"u}ck University, 49076 Osnabr{\"u}ck, Germany}

\author{T.~Schemme}
\affiliation{Department of Physics and Center of Physics and Chemistry of New Materials, Osnabr{\"u}ck University, 49076 Osnabr{\"u}ck, Germany}

\author{J.~Schmalhorst}
\affiliation{Center for Spinelectronic Materials and Devices, Physics Department, Bielefeld University, Universit{\"a}tsstra{\ss}e 25, 33615 Bielefeld, Germany}

\author{A.~Thomas}
\affiliation{Center for Spinelectronic Materials and Devices, Physics Department, Bielefeld University, Universit{\"a}tsstra{\ss}e 25, 33615 Bielefeld, Germany}
\affiliation{Leibniz Institute for Solid State and Materials Research Dresden (IFW Dresden), Institute for Metallic Materials, Helmholtzstrasse 20, 01069 Dresden, Germany}

\author{E.~Arenholz}
\affiliation{Advanced Light Source, Lawrence Berkeley National Laboratory, California 94720, USA}

\author{M.~Gorgoi}
\affiliation{Helmholtz-Zentrum f{\"u}r Materialien und Energie GmbH, 12489 Berlin, Germany}

\author{R.~Ovsyannikov}
\affiliation{Advanced Light Source, Lawrence Berkeley National Laboratory, California 94720, USA}

\author{S.~Bartkowski}
\affiliation{Department of Physics and Center of Physics and Chemistry of New Materials, Osnabr{\"u}ck University, 49076 Osnabr{\"u}ck, Germany}

\author{G.~Reiss}
\affiliation{Center for Spinelectronic Materials and Devices, Physics Department, Bielefeld University, Universit{\"a}tsstra{\ss}e 25, 33615 Bielefeld, Germany}

\author{J.~Wollschl{\"a}ger}
\affiliation{Department of Physics and Center of Physics and Chemistry of New Materials, Osnabr{\"u}ck University, 49076 Osnabr{\"u}ck, Germany}

\date{\today}% It is always \today, today,

\begin{abstract}
We study the underlying chemical, electronic and magnetic properties of a number of magnetite based thin films. The main focus is placed onto NiO/Fe$_3$O$_4$(001) bilayers grown on MgO(001) and Nb-SrTiO$_3$(001) substrates. We compare the results with those obtained on pure Fe$_3$O$_4$(001) thin films. It is found that the magnetite layers are oxidized and Fe$^{3+}$ dominates at the surfaces due to maghemite ($\gamma$-Fe$_2$O$_3$) formation, which decreases with increasing magnetite layer thickness. From a layer thickness of around 20 nm on the cationic distribution is close to that of stoichiometric Fe$_3$O$_4$. At the interface between NiO and Fe$_3$O$_4$ we find the Ni to be in a divalent valence state, with unambiguous spectral features in the Ni 2p core level x-ray photoelectron spectra typical for NiO. The formation of a significant NiFe$_2$O$_4$ interlayer can be excluded by means of XMCD. Magneto optical Kerr effect measurements reveal significant higher coercive fields compared to magnetite thin films grown on MgO(001), and a 45$^{\circ}$ rotated magnetic easy axis. We discuss the spin magnetic moments of the magnetite layers and find that the moment increases with increasing thin film thickness. At low thickness the NiO/Fe$_3$O$_4$ films grown on Nb-SrTiO$_3$ exhibits a significantly decreased spin magnetic moments. A thickness of 20 nm or above leads to spin magnetic moments close to that of bulk magnetite.
\end{abstract}

o% PACS, the Physics and Astronomy



\maketitle

% Introduction
\section{Introduction}\label{intro}

Transition metal oxides display a remarkable variety of properties as function of the complex interplay between the electron charge, spin and orbital degree of freedom. Among many other transition metal oxides as for instance perovskite based manganites, ferrites or cuprates, the oldest known magnetic material, namely magnetite (Fe$_3$O$_4$) is of special interest in current condensed matter and thin film physics. This fact stems partly from fundamental aspects as the high Curie temperature (860 K) for bulk material with a magnetic saturation moment of 4.07 $\mu_B$ per formula unit,\cite{wei29,sli80} the Verwey transition taking place at around 120 K,\cite{ver39,sen12} as well as the predicted 100\% spin polarization of Fe$_3$O$_4$,\cite{kat08} and partly this material is of special interest for various applications in medicine, catalysis and in particular for future spintronic devices such as spin valves or magnetic tunnel junctions.\cite{zut04,dre09,bli14,byr15} Recent approaches are pointing towards the synthesis of so-called all-oxide devices, e.g.\ by epitaxial growth of Fe$_3$O$_4$ on SrTiO$_3$(001) substrates to obtain the desired magnetite properties for further incorporation into Fe$_3$O$_4$/La$_{0.7}$Sr$_{0.3}$MnO$_3$ spin valves.\cite{moy15} Another promising route to build  a full oxide spin valve is to employ exchange bias between a ferro(i)magnetic and an antiferromagnetic oxide to manipulate the magnetization state of magnetite.\cite{wu13} Such an exchange interaction can be realized by coupling of Fe$_3$O$_4$ with antiferromagnetic NiO in thin film heterostructures.\cite{ber99,gat05,wu13} However, for the optimization of such potential all-oxidic devices as mentioned above a detailed characterization and understanding of the underlying structural, chemical, electronic and magnetic properties of the thin films and the interfaces, e.g.\ between magnetite and the substrate or the antiferromagnetic NiO is indispensable.\par

\begin{table}
\caption{Sample abbreviations (used in the text for brevity), and thin film thicknesses as determined from X-ray reflectivity (XRR) experiments.\label{table1}}
\begin{ruledtabular}
\begin{tabular}{@{}lb{10pt}rclrclr@{}}
 sample && substrate & $\vert$ & \multicolumn{4}{l}{thin film layer(s)} \\ [2pt]
\hline \\ [-8pt]
\textbf{S1}	 & & SrTiO$_{3}$ & $\vert$ & Fe$_{3}$O$_{4}$ & (12.3 nm) & & & \\
\textbf{S2}	 & & SrTiO$_{3}$ & $\vert$ & Fe$_{3}$O$_{4}$ & (21.5 nm) & & & \\
\textbf{S3}	 & & SrTiO$_{3}$ & $\vert$ & Fe$_{3}$O$_{4}$ & (33.0 nm) & & & \\ [3pt]
\textbf{SN1} & & SrTiO$_{3}$ & $\vert$ & NiO & (5.0 nm)  & $\vert$ & Fe$_{3}$O$_{4}$ & (6.2 nm) \\
\textbf{SN2} & & SrTiO$_{3}$ & $\vert$ & NiO & (9.7 nm)  & $\vert$ & Fe$_{3}$O$_{4}$ & (9.6 nm) \\
\textbf{SN3} & & SrTiO$_{3}$ & $\vert$ & NiO & (10.3 nm) & $\vert$ & Fe$_{3}$O$_{4}$ & (20.7 nm) \\ [3pt]
\textbf{M1}	 & & MgO & $\vert$ & Fe$_{3}$O$_{4}$ & (20.0 nm) & & & \\ [3pt]
\textbf{MN1} & & MgO & $\vert$ & NiO & (5.2 nm) & $\vert$ & Fe$_{3}$O$_{4}$ & (6.1 nm) \\
\textbf{MN2} & & MgO & $\vert$ & NiO & (11.5 nm) & $\vert$ & Fe$_{3}$O$_{4}$ & (10.5 nm) \\
\textbf{MN3} & & MgO & $\vert$ & NiO & (8.2 nm) & $\vert$ & Fe$_{3}$O$_{4}$ & (21.5 nm)
\end{tabular}
\end{ruledtabular}
\end{table}

%\begin{table}[hbt]
%	\centering
%	\begin{tabular}{|c|c|c|c|}
%        \hline
%		sample & substrate & NiO thickness (nm) & Fe$_3$O$_4$ thickness (nm) \\\hline
%		\textbf{S1} & SrTiO$_3$ & & 12.3 \\
%		\textbf{S2} & SrTiO$_3$ & & 21.5 \\
%		\textbf{S3} & SrTiO$_3$ & & 33.0 \\\hline
%       \textbf{SN1} & SrTiO$_3$ & 5.0 & 6.2 \\
%   	\textbf{SN2} & SrTiO$_3$ & 9.7 & 9.6 \\
%		\textbf{SN3} & SrTiO$_3$ & 10.3 & 20.7 \\\hline
%       \textbf{M1} & MgO & & 20.0 \\\hline
%        \textbf{MN1} & MgO & 5.2 & 6.1 \\
%		\textbf{MN2} & MgO & 11.9 & 10.5 \\
%		\textbf{MN3} & MgO & 8.2 & 21.5 \\\hline
%	\end{tabular}
%	\caption{Sample abbreviations (used in the text for brevity), and thin film thicknesses as determined from X-ray reflection (XRR) experiments.}
%	\label{table1}
%\end{table}

The lattice mismatch between magnetite (bulk lattice constant 0.83963 nm) and the doubled MgO bulk lattice constant (0.42117 nm) is only 0.3\%, making MgO an optimal candidate for epitaxial magnetite growth, which has been demonstrated many times using various deposition techniques.\cite{bal04,ste07,ber12,ber13,moy15b,sch15b} There are quite a few works on NiO/Fe$_3$O$_4$ bilayers and multilayers, mostly grown on MgO, investigating the structural properties of the heterostructures and the nature of exchange interaction between magnetite and (the antiferromagnetic) nickel oxide.\cite{bor95,kel02,gat05,sch15}\par
Thin magnetite films grown on SrTiO$_3$(001) allow tuning the conductivity via Nb-doping, which is of interest for some important applications. Quite a few studies reveal altered magnetic properties such as a strongly increased coercive field compared to films grown on MgO.\cite{che08b,wei10} Furthermore, Monti \emph{et al.}\ recently report that the magnetic easy axis points along the in-plane $\langle$100$\rangle$ film directions\cite{mon13} rather than the $\langle$110$\rangle$ directions for the easy axis mostly reported for SrTiO$_3$\cite{kal01} and MgO.\cite{bra08,sch15b} Despite the large lattice mismatch of -7.5\% between the doubled SrTiO$_3$ bulk lattice constant (0.3905 nm) and magnetite at least some of the thin films appear to grow almost relaxed on the SrTiO$_3$(001) surface,\cite{mon13,rub15} hence the reason of the altered magnetic properties including the role of epitaxial strain and potential formation of anti-phase domain boundaries (APB's) is still under discussion. As to coupled Fe$_3$O$_4$/NiO bilayers grown on SrTiO$_3$, up to now only Pilard \emph{et al.}\ report on the magnetic properties of the NiO/Fe$_3$O$_4$ interface.\cite{pil07} They find evidence for an 1.5 nm thick NiFe$_2$O$_4$ interfacial layer. An enhanced interfacial magnetization is also found by Krug \emph{et al.}, they present a rather complete study of the coupling on the NiO/Fe$_3$O$_4$(110) system.\cite{kru08}\par
Since the orientation of the interface is of utmost importance for the magnetic properties in oxidic thin film systems we want to perform a comprehensive study of NiO/Fe$_3$O$_4$(001) bi-layers grown by reactive molecular beam epitaxy (RMBE) on MgO(001) and Nb-doped SrTiO$_3$(001) substrates, respectively. We investigate the electronic structure and the chemical composition by means of depth selective hard x-ray photoelectron spectroscopy (HAXPES) and x-ray magnetic circular dichroism (XMCD) at the Fe L$_{2,3}$ and Ni L$_{2,3}$-edges. This complementary x-ray spectroscopic approach, along with magneto optical Kerr effect (MOKE) measurements allows us to develop a detailed picture of the overall electronic and magnetic structure of the magnetite based bi-layers in question. We compare our results with those obtained on pure magnetite thin films grown under equivalent conditions on Nb-SrTiO$_3$(001) and recent investigations reported in the literature.

\section{Experimental and Theoretical details}\label{exp}

Preparation of the NiO/Fe$_3$O$_4$ heterostructures, and in situ characterization by means of LEED and XPS have been performed in a multichamber ultra-high vacuum system. The XPS system is equipped with a Phoibos HSA 150 hemispherical analyzer and an Al K$\alpha$ anode (1486.6 eV), leading to an information depth ($\sim$3 nm mean free path) of about 5 nm in case of magnetite. Prior to thin film deposition the MgO(001) and SrTiO$_3$ (doped with 0.05 wt.\% Nb) substrates were cleaned by heating them up to 400$^{\circ}$C at an oxygen partial pressure of 1$\cdot 10^{-4}$ mbar for 1 hour. Afterwards the quality of the substrates has been checked by LEED. In a next step NiO layers were deposited under an oxygen partial pressure of 1$ \cdot 10^{-5}$ mbar at a substrate temperature of 250$^{\circ}$C (in order to prevent any Mg interdiffusion in case of MgO).\cite{kim09} After checking the quality of the NiO surfaces by LEED and XPS the magnetite layers have been grown on top of the NiO films at 250$^{\circ}$C and an oxygen partial pressure of 5$\cdot 10^{-6}$ mbar. Subsequently LEED and XPS have been performed in situ in order to check the magnetite surface reconstruction and chemical composition.
Then the samples were transported under ambient conditions for further characterization. MOKE measurements have been conducted with a longitudinal setup and a HeNe laser. The Kerr rotation has been measured with help of a photoelastic modulator (PEM). For the HAXPES experiments we have used the HIKE endstation of the KMC-1 beamline of the BESSY II synchrotron facility.\cite{gor09} Spectra have been recorded at photon energies ranging from
2.2 keV to 9.6 keV. The samples have been aligned at 3$^{\circ}$ grazing incidence as to the incoming x-ray beam. The Scienta R4000 photoelectron analyzer has been normal to the sample surface. All HAXPES has been taken at room temperature, as well as the XMCD spectra performed at the Fe L$_{2,3}$ and Ni L$_{2,3}$ edges at beamline 6.3.1 of the Advanced Light Source, Lawrence Berkeley Laboratory. We have utilized total electron yield (TEY) as detection mode, which is mainly sensitive to the surface near region. The external magnetic field of 1.5 T has been aligned parallel to the x-ray beam and has been reversed at each energy. The angle between sample surface and x-ray beam was 30$^{\circ}$. The degree of circular polarization has been about 55\%.\par
For the analysis of the Fe L$_{2,3}$ XMCD spectra, we have performed corresponding model calculations within the atomic multiplet and crystal field theory including charge transfer using the program \textsc{CTM4XAS}.\cite{dgr05,sta10}

%%%%%%%%%%%%%%%%%%%%%%%%%%%%%%%%%%%%%%%%%%%%%%%%%%%%%%%%%%%%%%%%%
%Results
\section{Results and Discussion}\label{results}
%%%%%%%%%%%%%%%%%%%%%%%%%%%%%%%%%%%%%%%%%%%%%%%%%%%%%%%%%%%%%%%%%

The Fe$_3$O$_4$/NiO-bilayers have been synthesized using the same setup and using the same growth conditions described very recently by Schemme \emph{et al.},\cite{sch15} see also the experimental section. All cleaned MgO(001) and SrTiO$_3$(001) substrates are checked by means of LEED for the expected (1 $\times$ 1) structure. Subsequently to each layer deposition, the NiO and magnetite layers are investigated for the typical NiO (1 $\times$ 1) LEED pattern and the characteristic ($\sqrt{2} \times \sqrt{2}R45^{\circ}$) superstructure for Fe$_3$O$_4$. Ni 2p and Fe 2p XPS core levels confirm the formation of stoichiometric NiO and Fe$_3$O$_4$ deposited on MgO(001) and SrTiO$_3$(001). These data are discussed in detail along with a comprehensive structural analysis elsewhere.\cite{kus15} Here we only sum up the results for the layer thicknesses (determined by X-ray reflectivity XRR) in table \ref{table1} in order to report on the thickness dependence of the electronic structure and magnetic properties.

\subsection{Electronic structure: Hard x-ray photoelectron spectroscopy (HAXPES)}

\begin{figure}[hbt]
    \centering
        \includegraphics[width=8.75cm]{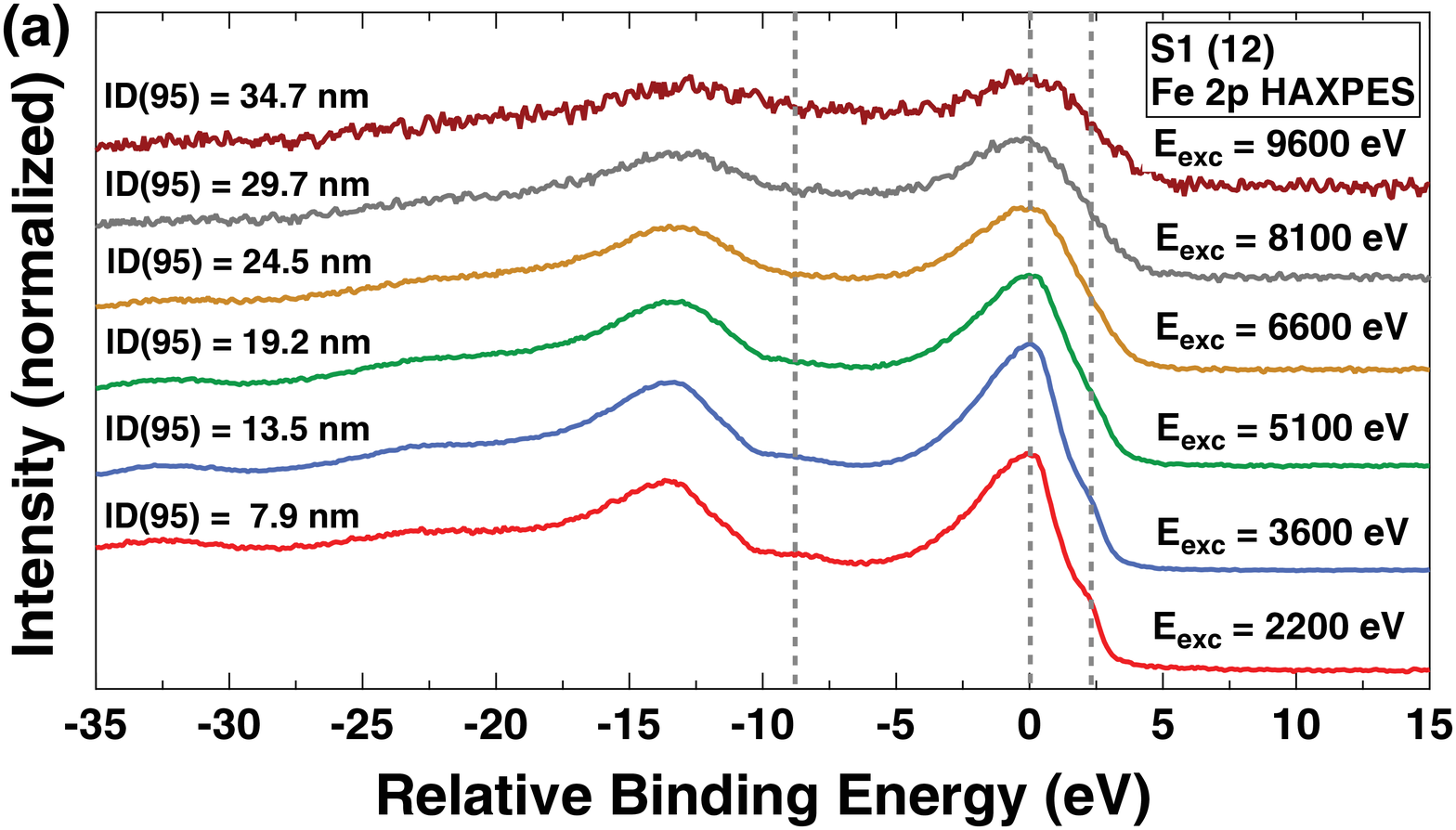}
        \includegraphics[width=8.75cm]{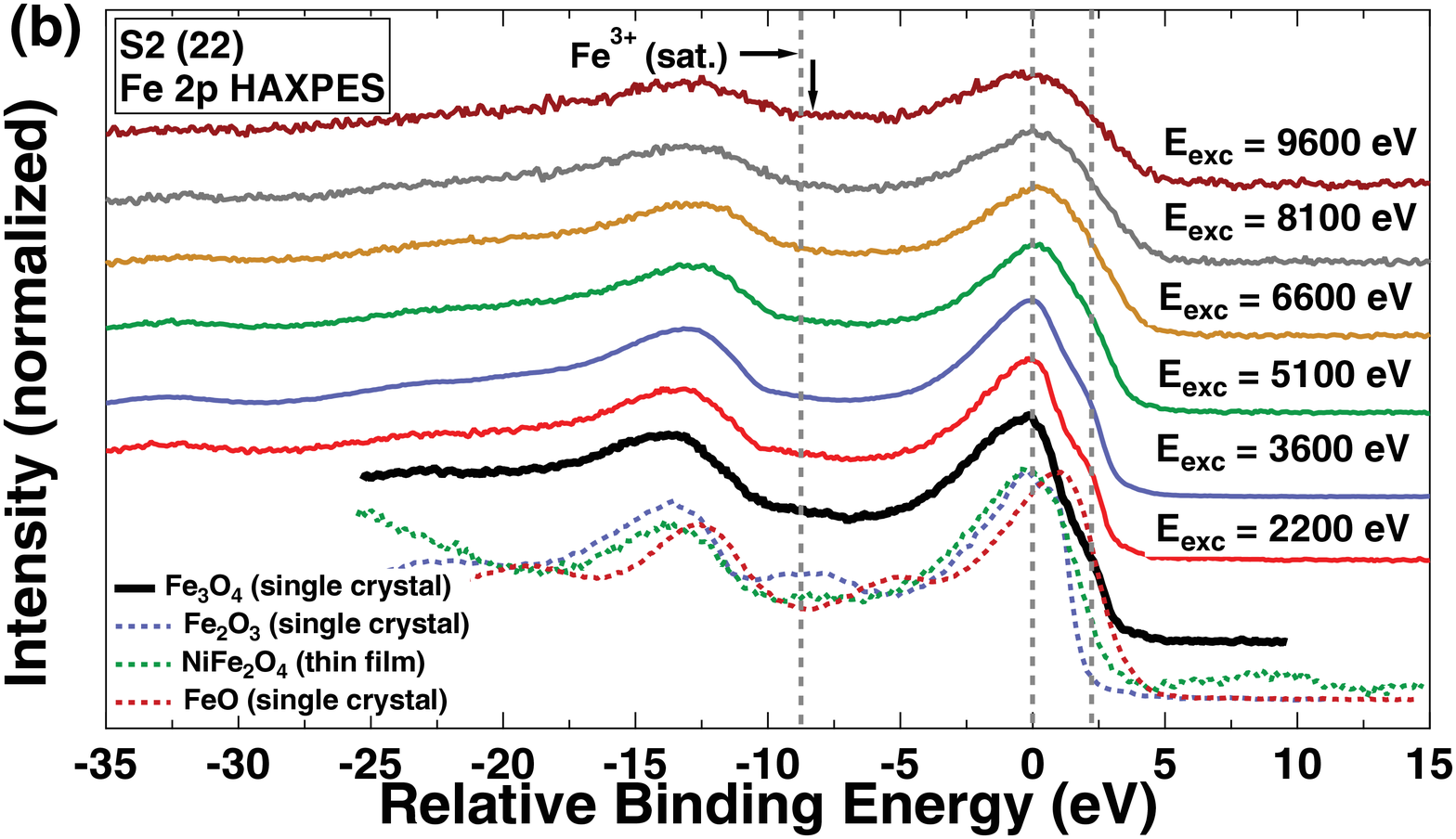}
        \includegraphics[width=8.75cm]{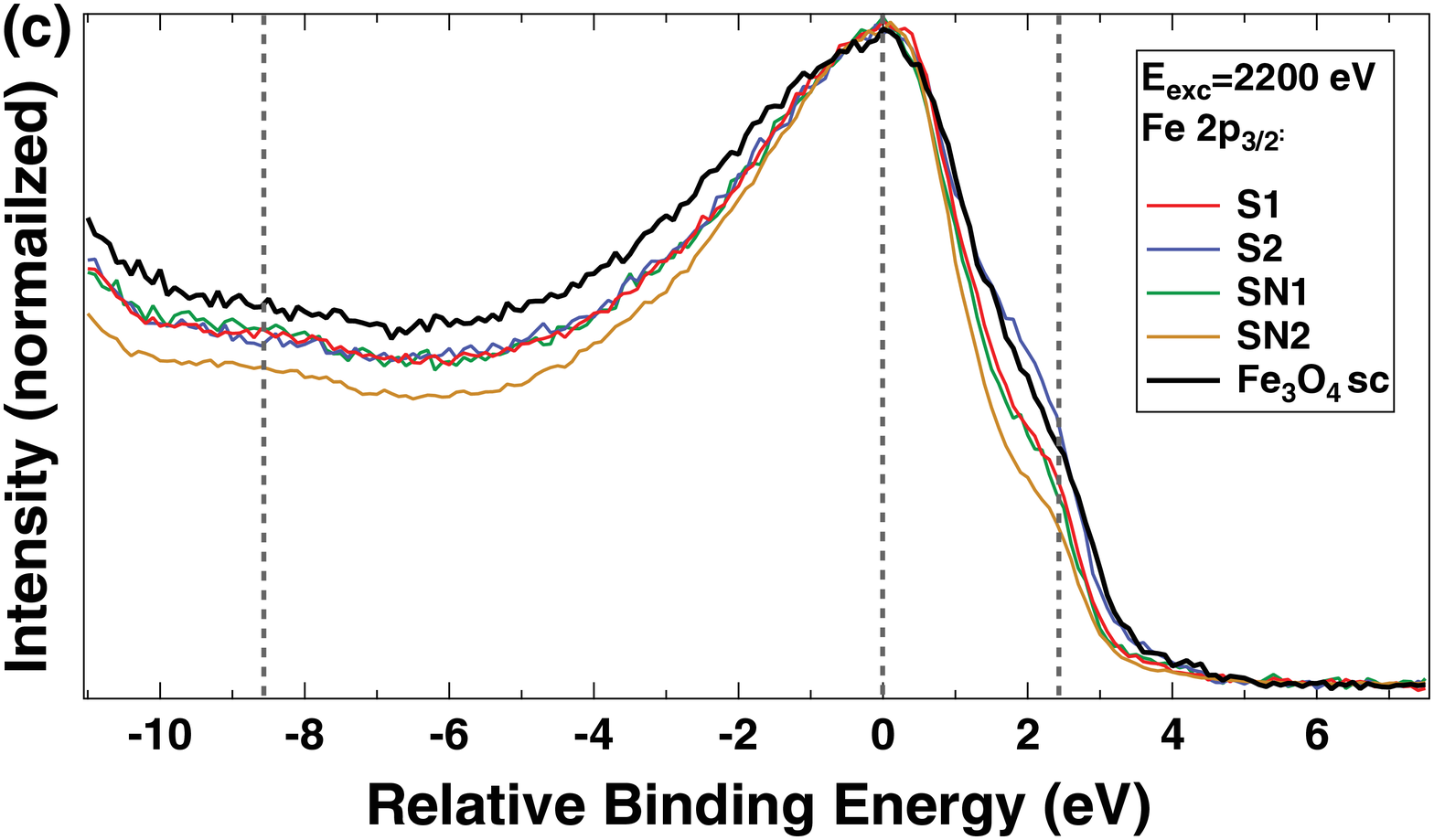}
    \caption{\label{Fig1} Fe 2p HAXPES of samples \textbf{S1} (a) and \textbf{S2} (b) recorded at excitation energies between 2200 eV and 9600 eV. Reference spectra of FeO, Fe$_3$O$_4$, NiFe$_2$O$_4$, and Fe$_2$O$_3$\cite{kle14} are also shown for comparison. (c) Fe 2p$_{3/2}$ core level of samples \textbf{S1}, \textbf{S2}, \textbf{SN1} and \textbf{SN2} recorded at E$_{exc}$ = 2200 eV.}
\end{figure}

\begin{figure}[hbt]
    \centering
        \includegraphics[width=8.25cm]{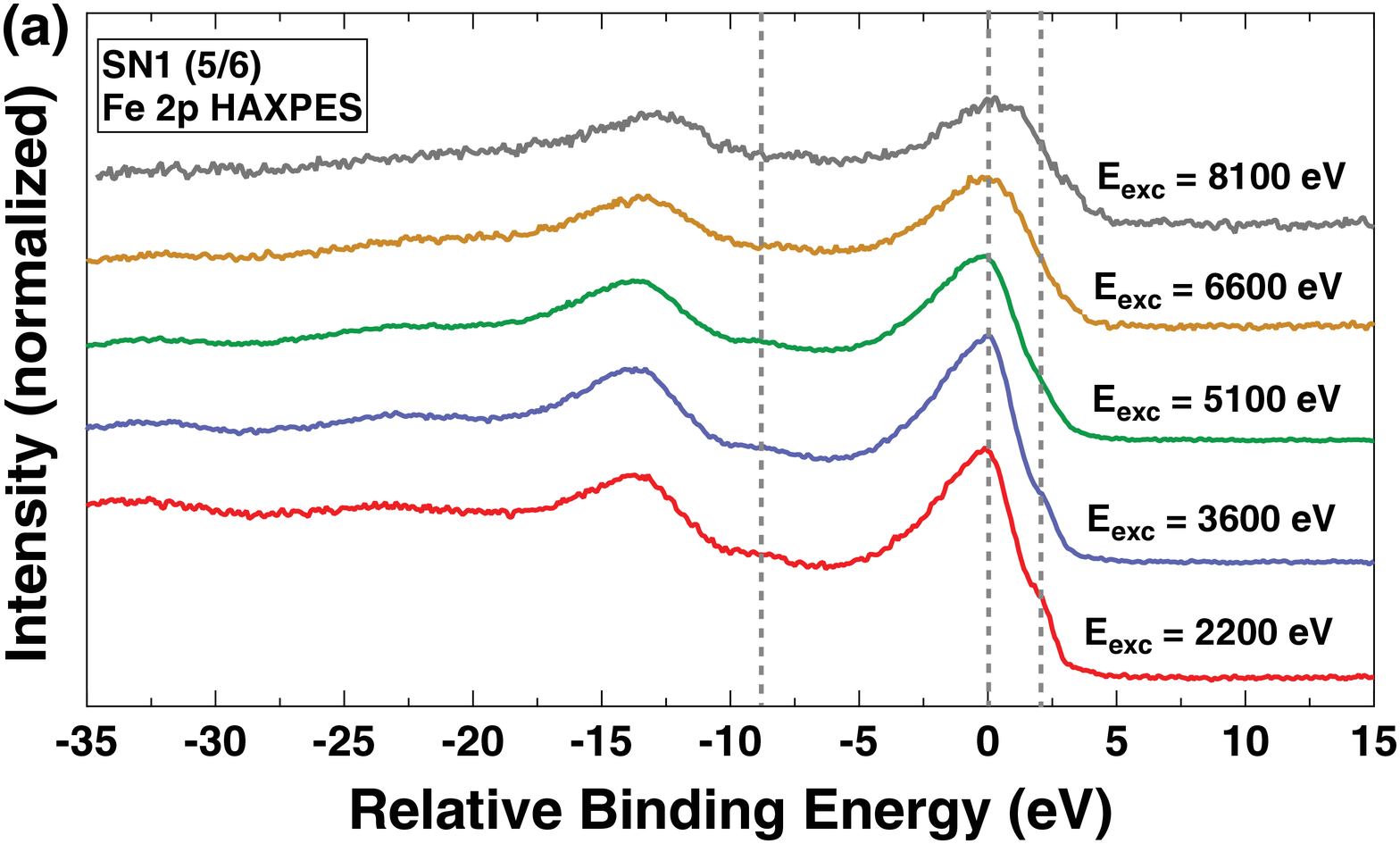}
        \includegraphics[width=8.25cm]{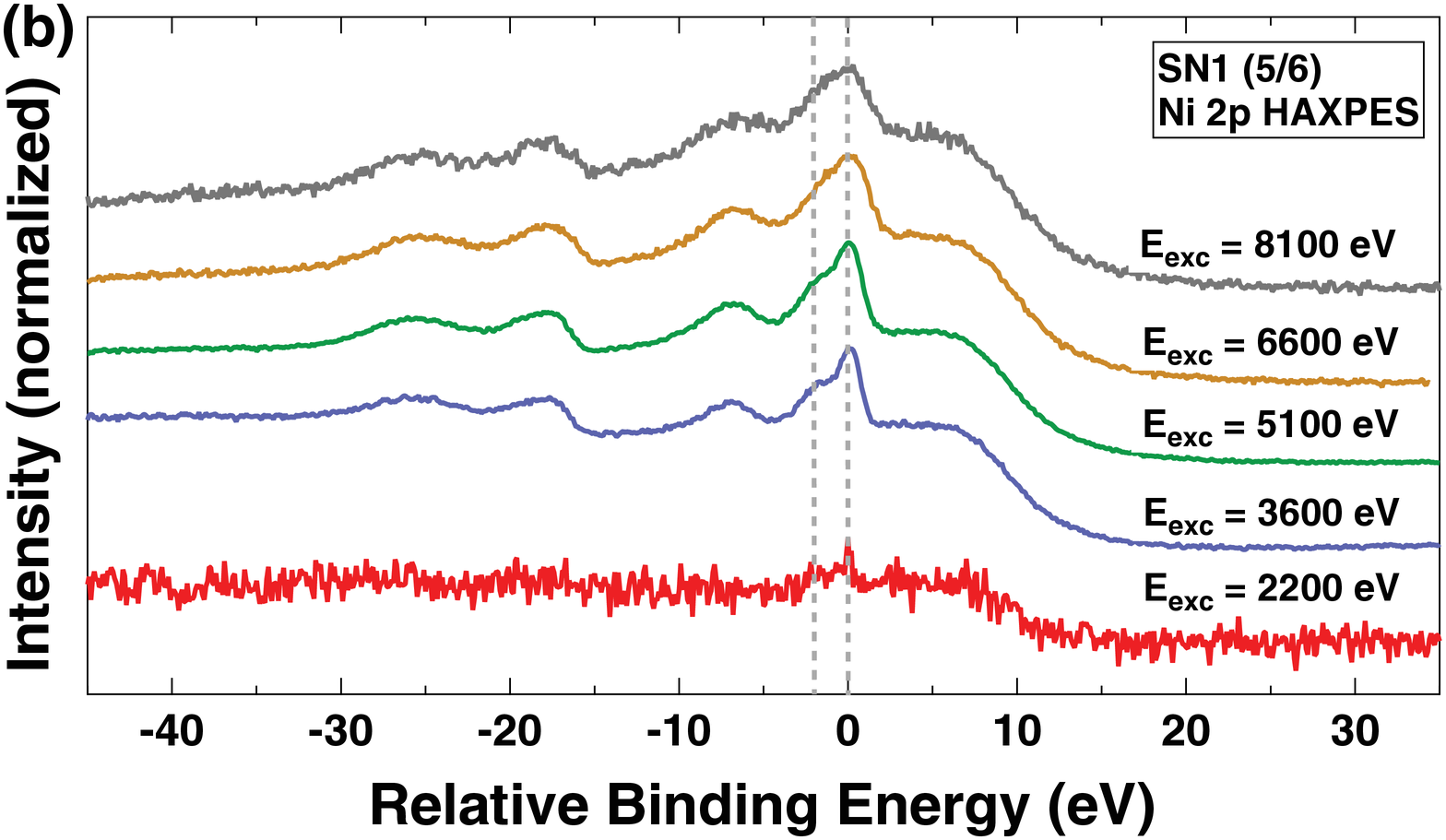}
        \includegraphics[width=8.25cm]{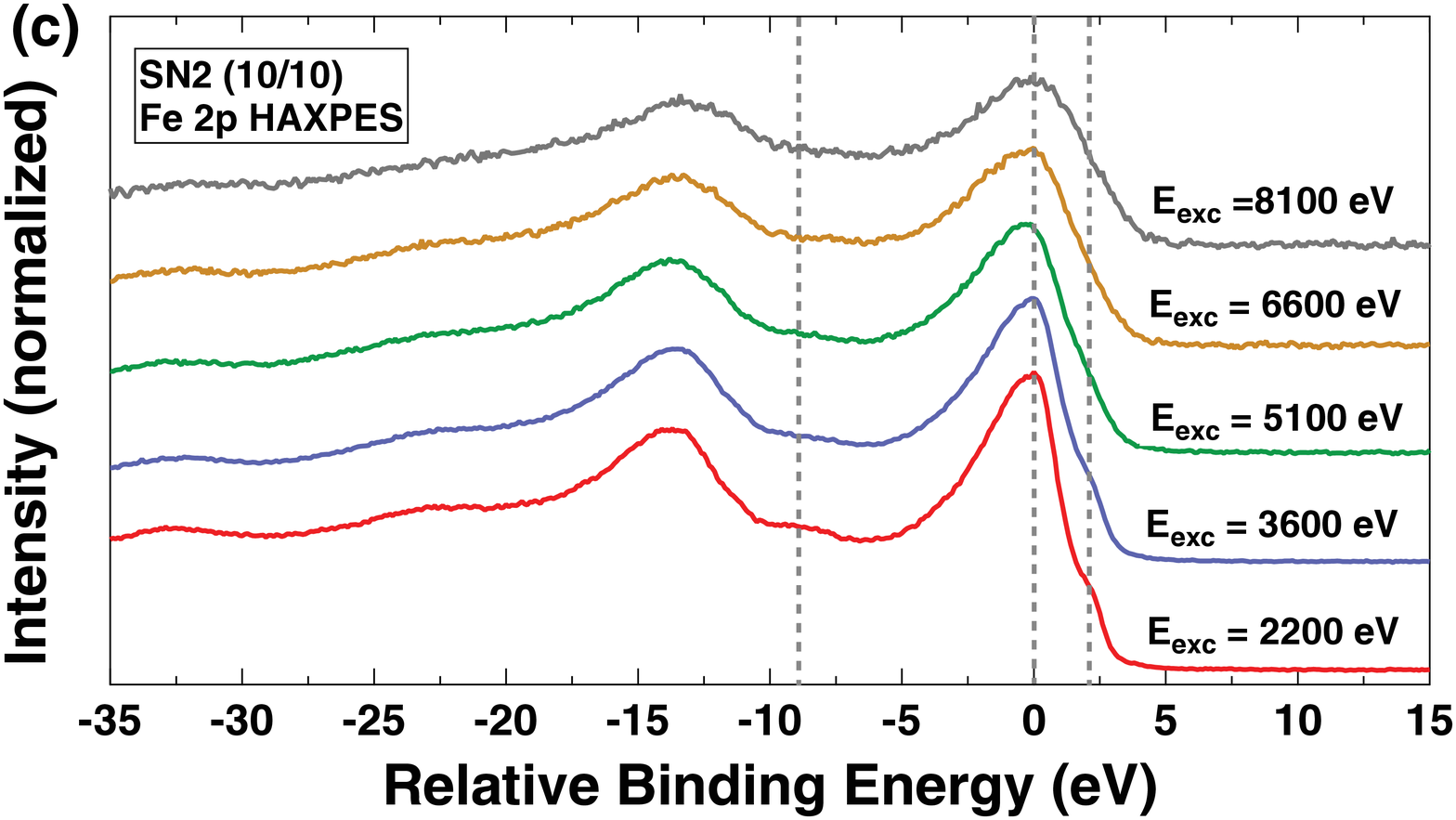}
        \includegraphics[width=8.25cm]{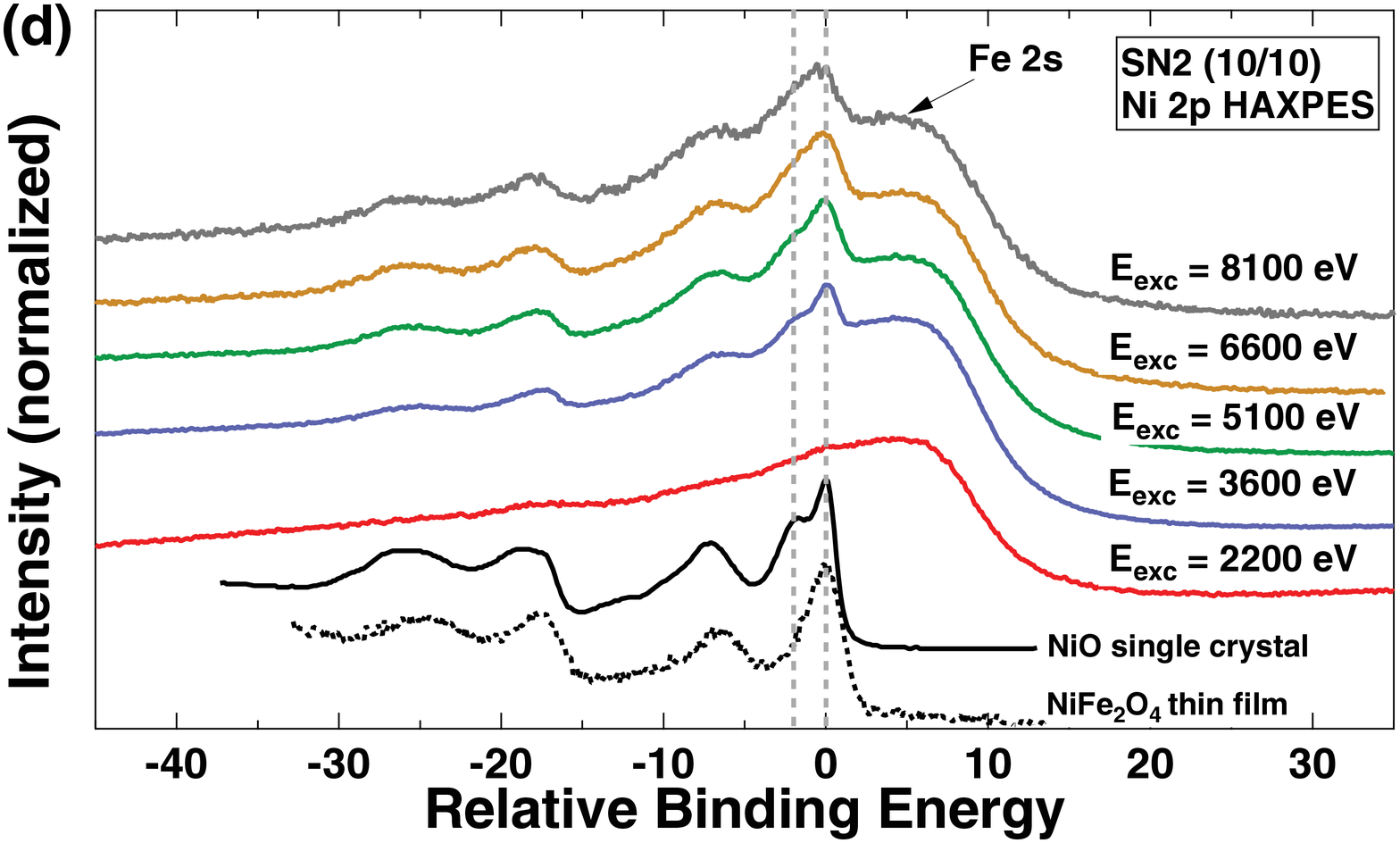}
    \caption{\label{Fig2} HAXPE-spectra of samples \textbf{SN1} and \textbf{SN2} recorded at excitation energies between 2200 eV and 8100 eV. (a) and (b) Fe 2p and Ni 2p HAXPES of sample \textbf{SN1}. (c) and (d) Fe 2p and Ni 2p HAXPES of sample \textbf{SN2}, reference spectra of NiO\cite{uhl92} and NiFe$_2$O$_4$\cite{kle14} are also shown for comparison.}
\end{figure}

\begin{figure}[hbt]
    \centering
        \includegraphics[width=8.0cm]{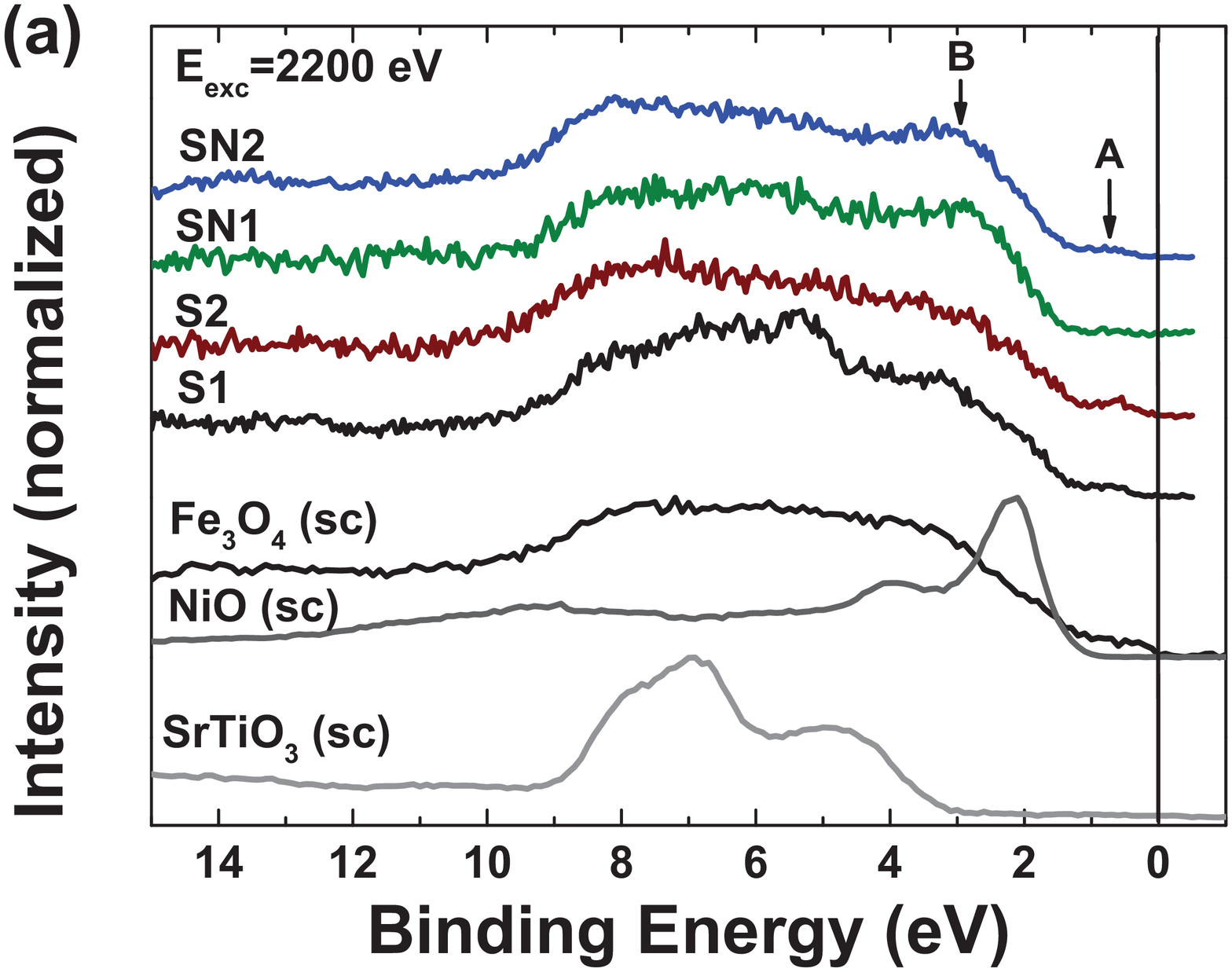}
        \includegraphics[width=8.0cm]{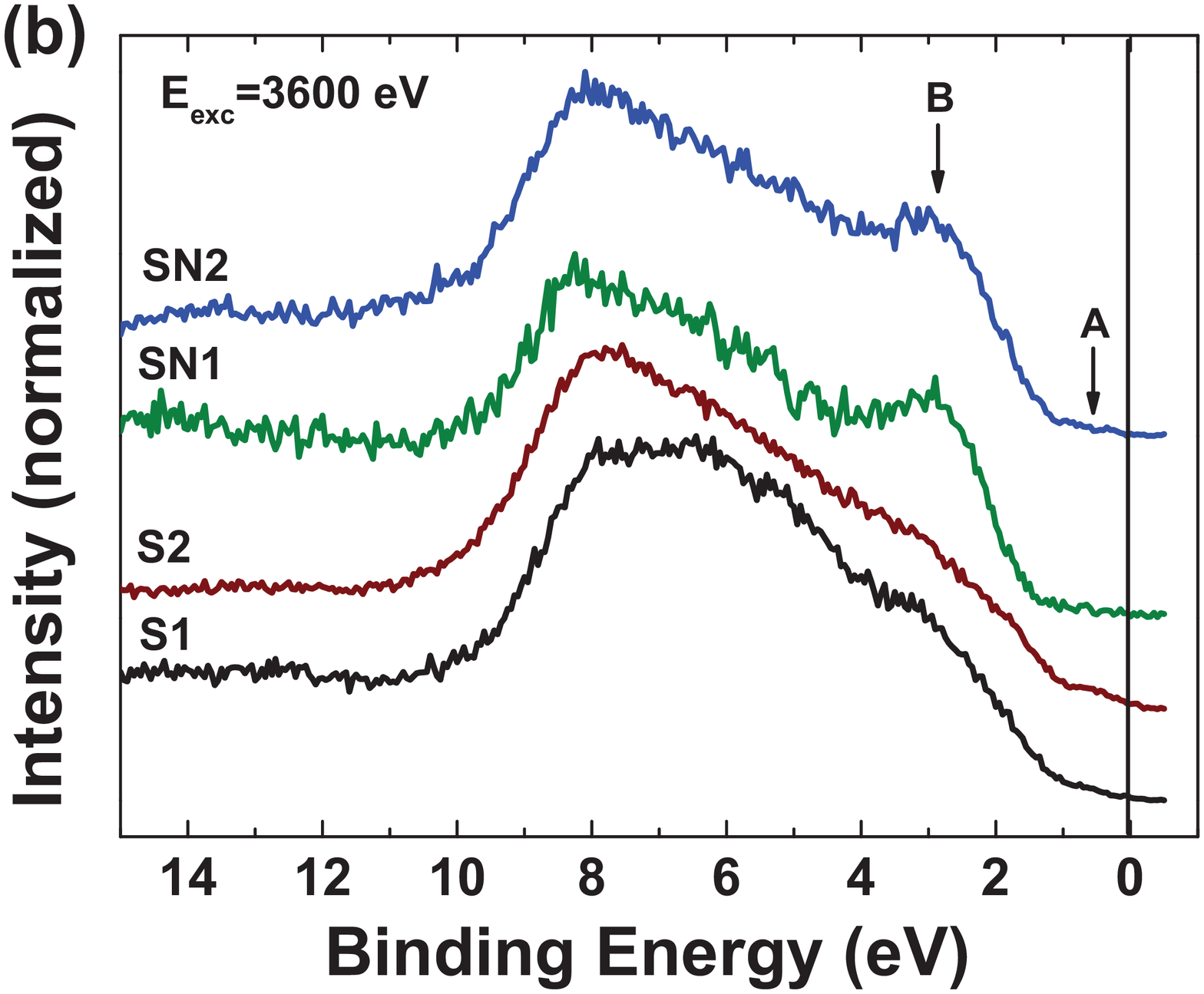}
    \caption{\label{Fig3} HAXPES valence bands of samples \textbf{S1}, \textbf{S2}, \textbf{SN1} and \textbf{SN2} recorded at excitation energies of 2200 eV (a) and 3600 eV (b). Reference spectra of SrTiO$_3$, NiO\cite{uhl92} and Fe$_3$O$_4$ are also shown for comparison.}
\end{figure}

In this section we want to clarify the chemical properties of the thin film surfaces and interfaces in detail. HAXPES measurements have been demonstrated to be a powerful tool for chemical depth profile and hence the characterization of buried layers and interfaces.\cite{clae09,hop15} All HAXPES measurements presented in Figs.~\ref{Fig1} to \ref{Fig3} are recorded in the geometry described in section \ref{exp}. The probing, or information depth (ID(95)) is defined as the depth from which 95\% of the photoelectrons of the spectra come from. ID(95) can be derived by the following equation:\cite{pow09} ID(95) = $\mathrm{{\lambda}_{IMFP}cos(\alpha)ln(1-95/100)}$.
Here $\alpha$ represents the off-normal emission angle. $\mathrm{\lambda_{IMFP}}$ is the inelastic mean free path of the photoelectrons. We derive $\mathrm{\lambda_{IMFP}}$ for the Fe$_3$O$_4$  top layers of the thin films and bilayers by employing the TTP-2 equation\cite{tan88,tan05} and subsequently use the equation for calculating ID(95) to estimate the overall information depth. The resulting information depth for magnetite are denoted in Fig.~\ref{Fig1}a for each excitation energy.

Fig.~\ref{Fig1} displays the Fe 2p spectra performed at excitation energies between 2200 eV and 9600 eV of samples \textbf{S1} and \textbf{S2}. The spectra have been normalized and re-scaled to the Fe 2p$_{3/2}$ maximum, so comparison between the spectra is straightforward. For stoichiometric Fe$_3$O$_4$ one expects the Fe 2p$_{3/2}$ binding energy at around 710.6 eV (0 eV on the relative binding energy scale chosen here), and a structureless region between the Fe 2p$_{3/2}$ and the Fe 2p$_{1/2}$ peaks without satellite peak as previously reported,\cite{yam08} since the charge transfer satellites add up in such a way for a mixed valence state found in Fe$_3$O$_4$. As to sample \textbf{S1} (12 nm) (Fig.~\ref{Fig1}a) there appears to be an excess of trivalent ions observed by means of the more surface sensitive 2200 eV photon energy. For E$_{exc}$ = 2200 eV the typical Fe$^{3+}$ charge transfer satellite is also visible at  8.5 eV  above the Fe 2p$_{3/2}$ maximum (see also gray vertical line, corresponding to a binding energy of around 719 eV). However, also a pronounced shoulder located around 2.2 eV at the low binding energy side (corresponding to 708.4 eV binding energy) is present indicating the presence of a significant amount of divalent iron ions.\cite{bie11} We find no indications of metallic iron which should manifest itself at 707 eV and thus 3.6 eV below the Fe 2p$_{3/2}$ maximum on the relative binding energy scale chosen here. At higher excitation energies the Fe$^{3+}$ charge transfer satellite is not visible anymore and, the overall Fe 2p spectral shape indicates a stoichiometric 2:1 ratio of Fe${^{3+}}$ to Fe$^{2+}$ ratio.\par
For the thicker Fe$_3$O$_4$ film \textbf{S2} (22 nm) the excess of Fe$^{3+}$ in the surface near layers is significantly less pronounced in comparison with sample \textbf{S1}. The Fe$^{3+}$ charge transfer satellite is less pronounced, the feature stemming from Fe$^{2+}$ is more pronounced in the E$_{exc}$=2200 eV spectrum of \textbf{S2} as compared to that of \textbf{S1} (see also Fig.~\ref{Fig1}c). At higher excitation energies (5100 eV and above) the Fe 2p spectra of \textbf{S1} resemble that of magnetite. The Fe$^{2+}$ characteristic low binding energy shoulder becomes less visible due to the limited resolution of the spectra at higher excitation energies (E$_{exc}$=6600 eV and above). It is also noteworthy that the Fe 2p spectra of sample \textbf{S2} taken at 8100 eV and 9600 eV also exhibit a weak Fe$^{3+}$ charge transfer satellite indicating a potential slight excess of Fe$^{3+}$ close to or at the interface between the magnetite thin film and the SrTiO$_3$. No such features are visible in the corresponding spectra of sample \textbf{S1}.\par

Fig.~\ref{Fig2} depicts the Fe 2p and Ni 2p HAXPES of samples \textbf{SN1} and \textbf{SN2}. Similar to samples \textbf{S1} and \textbf{S2} the Fe 2p spectra show a weak Fe$^{3+}$ charge transfer satellite at E$_{exc}$ = 2200 eV (Figs.~\ref{Fig2}a and c). At higher excitation energies the spectra are identical to that of Fe$_3$O$_4$, despite in particular for sample \textbf{SN1} a weak Fe$^{3+}$ charge transfer satellite is also present at higher excitation energies indicating a small excess of trivalent ions at the interface between Fe$_3$O$_4$ and NiO.\cite{uhl92} The corresponding Ni 2p spectra are overlapped by broad Fe 2s states (Figs.~\ref{Fig2}b and d). The overall shape of the Ni 2p spectra corresponds perfectly to that of NiO. Also the binding energies of the Ni 2p$_{3/2}$ and Ni 2p$_{1/2}$ main lines and the corresponding charge transfer satellites are typical for Ni$^{2+}$. Furthermore all Ni 2p spectra exhibit a prominent feature at the high binding energy side. The shoulder located 2 eV above the Ni 2p$_{3/2}$ maximum (indicated by vertical lines in Figs.~\ref{Fig2}b and d) is a specific feature of bulk NiO,\cite{ald96} which has been associated with a non-local screening effect. The double structure is clearly visible in all Ni 2p spectra, even in the E$_{exc}$ = 2200 eV spectrum of sample \textbf{SN1} despite the rather poor signal to noise ratio. For sample \textbf{SN2} there is only the rather broad Fe 2s peak visible as the information depth is too low to probe the NiO layer at excitation energy 2200 eV.\par

The Fe 2p and Ni 2p HAXPES results confirm that stoichiometric Fe$_3$O$_4$ layers are formed on all samples investigated by HAXPES. Only small traces of Fe$^{3+}$ excess might be found at the interface between Fe$_3$O$_4$ and SrTiO$_3$ of sample \textbf{S2}, and the Fe$_3$O$_4$/NiO interface of sample \textbf{SN1}. At the surface of all magnetite top layers a certain degree of Fe$^{3+}$ excess is found indicating a potential maghemite ($\gamma$-Fe$_2$O$_3$) formation at the surface. It seems that thinner magnetite layers tend to enhanced $\gamma$-Fe$_2$O$_3$ formation or, respectively, to an increased amount of trivalent iron ions in the surface near layers. All Ni 2p spectra are, independent of the excitation energy, a quite perfect fingerprint of bulk NiO within the limits of this experiment. Hence, the formation of a NiFe$_2$O$_4$ interface layer or considerable amounts of NiFe$_2$O$_4$ islands or clusters can be rather excluded. Moreover, no Fe or Ni metallic contributions are investigated in the HAXPES spectra.\par

Next we will discuss the occupied total densities of states (tDOS) along the measured XPS valence band spectra (Fig.~\ref{Fig3}) in order to learn details about the electronic structure at the Fe$_3$O$_4$/NiO interface.

Fig.~\ref{Fig3}a depicts the total densities of states of samples \textbf{S1}, \textbf{S2}, \textbf{SN1}, and \textbf{SN2} with those of Fe$_3$O$_4$, NiO, and SrTiO$_3$ single crystals. The latter spectra have been performed with a lab based monochromatized Al K$\alpha$ source (1486.6 eV). The valence band spectrum of sample \textbf{S2} is very similar to that of the Fe$_3$O$_4$ single crystal. For sample \textbf{S1} the region between 5-9 eV shows somewhat more intensity compared to single crystal Fe$_3$O$_4$, likely due to contributions from the SrTiO$_3$ substrate. The valence band spectra of the double layers \textbf{SN1} and \textbf{SN2} are also quite similar to that of magnetite. However, in particular in the valence band of sample \textbf{SN1} there are a few differences in detail. It appears that feature A is almost vanished in the valence band of sample \textbf{SN1}. As to theory\cite{ani96,fon05} this feature stems mainly from $t_{2g}$ spin down states of octahedrally coordinated Fe cations. This trend is confirmed by the spectra taken at E$_{exc}$ = 3600 eV (Fig.~\ref{Fig3}b). Whereas feature A becomes weaker or vanishes in the XPS valence bands of samples \textbf{SN1} and \textbf{SN2}, feature B, representing the (localized) Ni 3d states of NiO becomes stronger, confirming that also the interface and the underlying NiO layer is probed. The overall shape of the valence band spectra becomes different at excitation energy 3600 eV which is due to the different atomic cross sections at higher energies.\par

The valence band spectra presented in Fig.~\ref{Fig3} recorded at E$_{exc}$ = 2200 eV and 3600 eV can be understood as being a superposition of the electronic density of states of the Fe$_3$O$_4$ thin films including the interfaces and partly the NiO buffer layers or the SrTiO$_3$ substrates, respectively. Therefore we can not unambiguously discriminate between interface effects and size effects in the magnetite layer. We want to point out that a low density of states at the interface between Fe$_3$O$_4$ and NiO has been investigated before by growing ultrathin NiO layers onto an Fe$_3$O$_4$ single crystal and employing ultraviolet photoelectron spectroscopy (UPS).\cite{wan08} Another possible reason for the observed behavior is that thinner Fe$_3$O$_4$ films (e.g.~ 6 nm (sample \textbf{SN1})) are different compared to somewhat thicker magnetite layers as to their electrical properties.\cite{eer02} To further elucidate potential influences of different substrates or the presence of a buffer layer like NiO on the electronic structure near Fermi energy and hence the electrical properties further complementary experiments such as resistivity measurements are highly desirable.

\subsection{Magnetic properties: MOKE}

\begin{figure}[hbt]
    \centering
        \includegraphics[width=7.0cm]{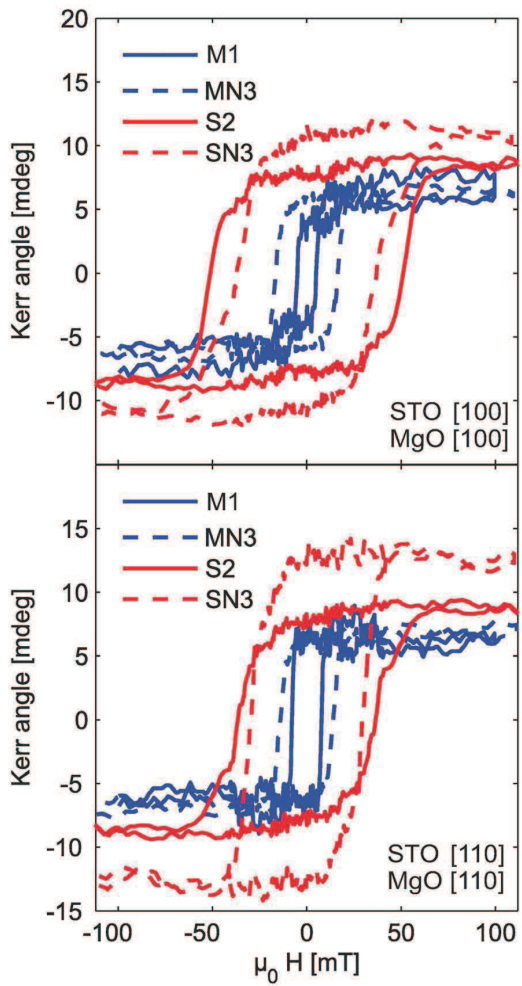}
    \caption{\label{Fig3B} Magnetization curves along the substrates [100] direction (upper panel) and the substrates [110] direction (lower panel) of samples \textbf{S2}, \textbf{SN3}, and \textbf{MN3}. The corresponding magnetization curves for a 20 nm Fe$_3$O$_4$ film grown on MgO(001) (\textbf{M1}) is also shown for comparison.\cite{sch15b}}
\end{figure}

Figure \ref{Fig3B} depicts the magnetization curves of the magnetic easy and hard directions of samples \textbf{S2}, \textbf{SN3} and \textbf{MN3} probed by MOKE. The results of a 20nm Fe$_3$O$_4$ film grown on MgO(001) are also plotted for comparison.\cite{sch15b}

Firstly, a significant increase of the coercive field is found compared to Fe$_3$O$_4$ films on MgO(001). This accounts in particular for the samples grown on Nb-SrTiO$_3$(001) substrates (\textbf{S2} and \textbf{SN3}). It is also noteworthy that a NiO buffer layer leads to an increase of the coercive field compared to corresponding Fe$_3$O$_4$ single layers in case of MgO(001) substrates but to a decrease of the coercive field in case of Nb-SrTiO$_3$(001) substrates. Whereas the latter (reference) sample the magnetic in-plane easy axis (high coercive field) are aligned along the magnetite $\langle$110$\rangle$ direction in well agreement with earlier reports,\cite{mar94,che08,sch15b} the other three samples seem to have $\langle$100$\rangle$ in-plane magnetic easy axis. The strong increase of the coercive field to around 50 mT in the easy magnetic direction as well as the rotated magnetic easy axis along the $\langle$110$\rangle$ in-plane directions have been recently also reported for thicker magnetite thin films (50-160 nm) grown by pulsed laser deposition on Nb-SrTiO$_3$(001).\cite{mon13}
Obviously, a NiO buffer layer between the magnetite layer and the substrate seems to lead to a 45$^{\circ}$ rotated magnetic easy in-plane axis independent of the MgO or SrTiO$_3$(001) substrate compared to Fe$_3$O$_4$/MgO(001). Potential reasons for this behavior concerning the magnetic easy axis rotation may stem from structural parameters such as interface structure and roughness or epitaxial strain,\cite{sch15c} or from potential presence of anti-phase domain boundaries (APBs) which may form during epitaxial thin film growth.\cite{mon13} A more complete study of the magnetic anisotropy, e.g.\ by means of angular resolved MOKE measurements and ferromagnetic resonance (FMR) experiments is ongoing and will be presented elsewhere.\cite{kus15} Moreover, despite the closer investigation of APB formation is an obstacle, further work to understand and analyse the role of APBs by means of high resolution TEM, or LEEM/PEEM experiments during thin film growth would be desirable for thin magnetite films and heterostructures.\cite{mar97,eer03,mcg08,mon13,mck14}

\subsection{Magnetic properties: X-ray magnetic circular dichroism (XMCD)}

\begin{figure}[hbt]
    \centering
        \includegraphics[width=7.75cm]{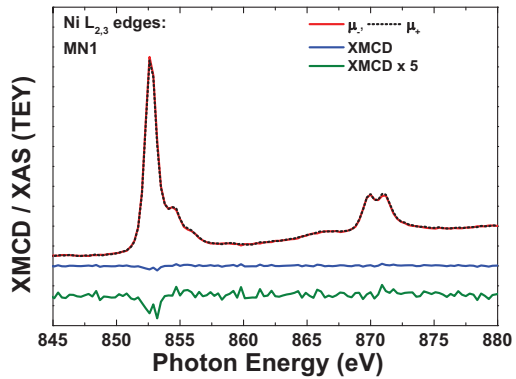}
    \caption{\label{Fig4} Polarization dependent X-ray absorption spectra performed at the Ni L$_{2,3}$ edges of samples \textbf{MN1} along with the resulting XMCD difference spectrum. The latter has been also multiplied by a factor of five for better visibility.}
\end{figure}

Fig.~\ref{Fig4} presents the Ni L$_{2,3}$-edge XA-spectra of sample \textbf{MN1} recorded at room temperature in an external magnetic field of $\pm$ 1.5T with circularly polarized light. The spectra show only a residual dichroic signal of  0.85\%. The TEY detection mode is interface sensitive due to the $1/e$ probing depth.\cite{ohl01} Furthermore, the Ni L$_{2,3}$-XMCD should be rather strong in case of a NiFe$_2$O$_4$ interlayer formation since in case of ultrathin films or spinel interfaces, a strongly enhanced magnetic moment due to proximity effects has been reported,\cite{nel10} and also found for ultrathin Fe$_3$O$_4$(110) films grown on NiO.\cite{kru08} Krug \emph{et al.}\ could even quantify the NiFe$_2$O$_4$ interlayer reconstruction to around one monolayer and deduced the Ni magnetic moments from the sum rules.\cite{kru08} In contrast, the very weak Ni L$_3$-XMCD, and within the experimental signal to noise ratio vanishing Ni L$_2$-XMCD (Fig.~\ref{Fig4}) indicate that there is no substantial formation of a NiFe$_2$O$_4$ interlayer or clusters at the interface between the Fe$_3$O$_4$(100) and the NiO layers studied in the present work. This result confirms the Ni 2p HAXPE-spectra of samples \textbf{SN1} and \textbf{SN2}, as the corresponding Ni 2p HAXPES results of \textbf{SN1} and \textbf{SN2} indicate charge transfer satellites typical for NiO.

\begin{figure}[hbt]
    \centering
        \includegraphics[width=8.125cm]{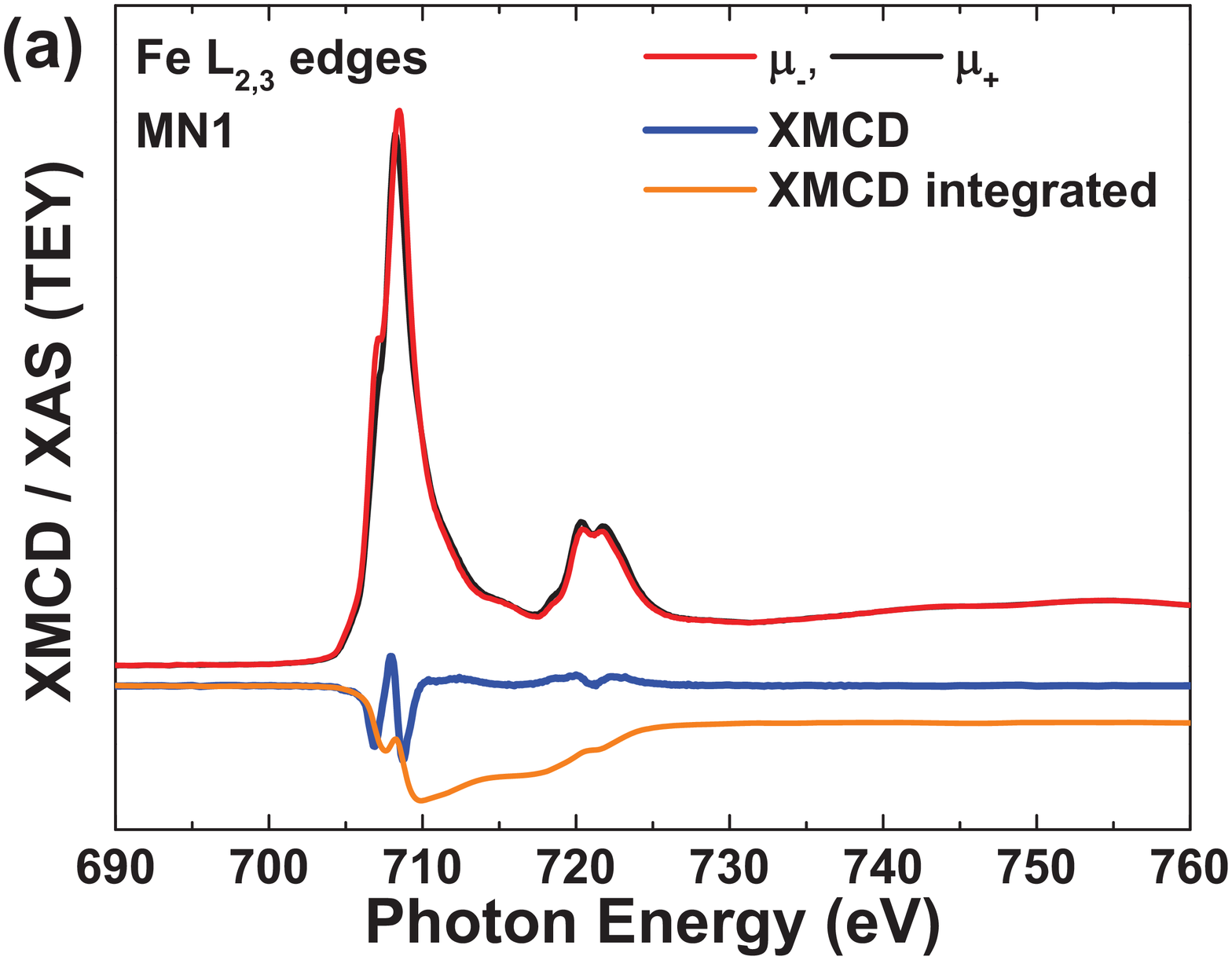}
        \includegraphics[width=8.125cm]{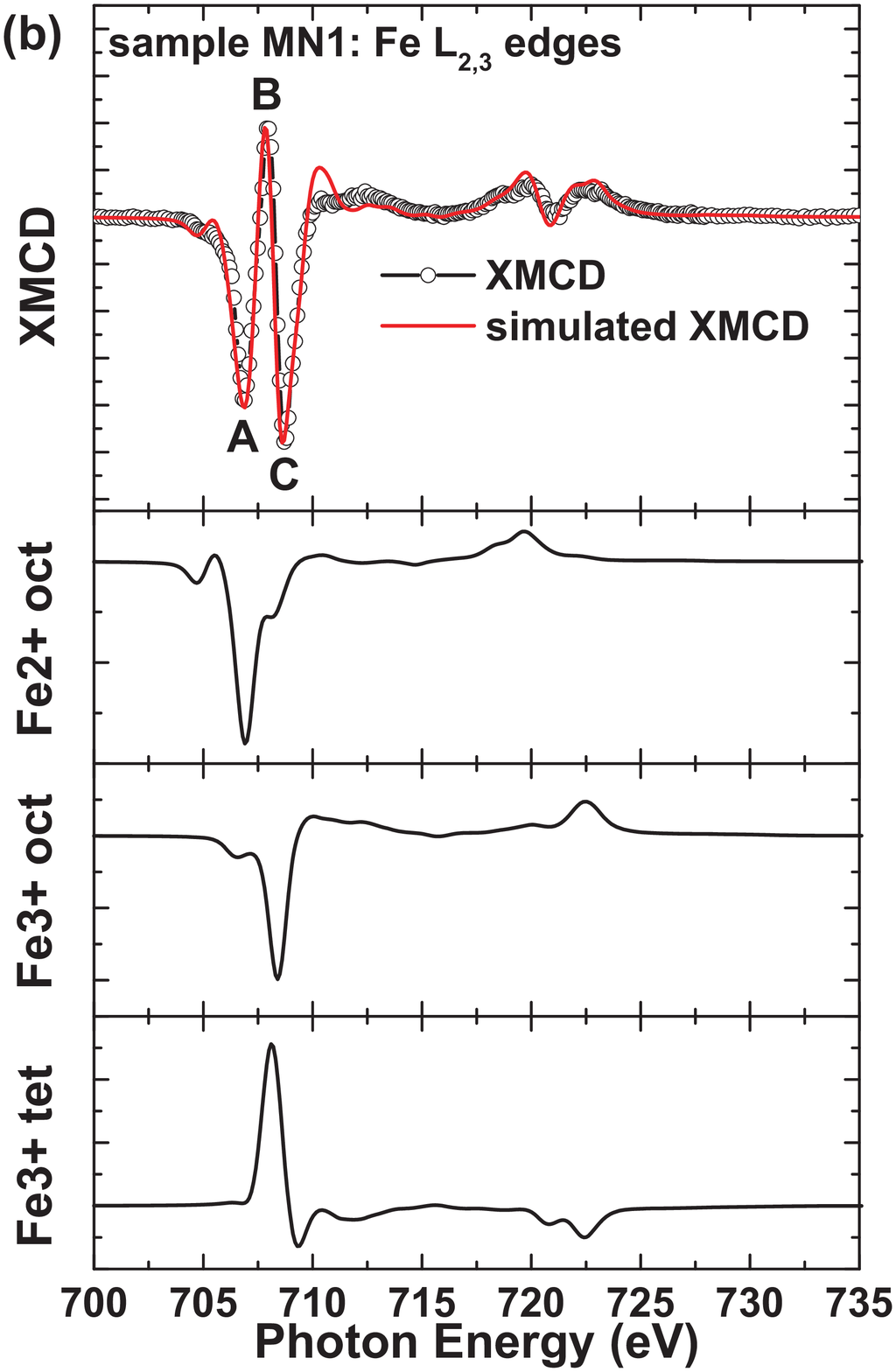}
    \caption{\label{Fig5} (a) Polarization dependent XA-spectra performed at the Fe L$_{2,3}$ edges of sample \textbf{MN1} along with the resulting XMCD and its integral. (b) XMCD difference spectra and corresponding simulated fit as a result from summarizing the suitable amounts of the charge transfer multiplet simulations of Fe$^{2+}$ ions in octahedral coordination, and Fe$^{3+}$ ions in octahedral and tetrahedral coordination, respectively.}
\end{figure}

Turning to the Fe L$_{2,3}$ edges, Fig.~\ref{Fig5}a depicts the XA-spectra taken at $\pm$ 1.5T of sample \textbf{MN1}, the resulting XMCD difference spectrum and the integral of the XMCD. The lower panel of Fig.~\ref{Fig5} shows the experimental XMCD of \textbf{MN1} along with the best fit obtained from an appropriate superposition of the single ion charge transfer multipet simulations plotted below.\par
The charge transfer multiplet spectra are calculated in the following way. In a first step, the Slater integrals and the spin – orbit couplings are calculated in spherical symmetry. The $d$--$d$ and $p$--$d$ integrals are reduced to 80\% of their atomic value in order to account for screening, whereas the spin-orbit parameters are not reduced. Then crystal fields of $10Dq$=1.0 eV (Fe$^{2+}_{oct}$, Fe$^{3+}_{oct}$), and -0.6 eV for the Fe$^{3+}_{tet}$ using the crystal field approach of Kuiper \emph{et al.}\cite{kui97} as starting point. An exchange field $g\mu_{B}$=0.03 eV is also applied. In a last step, charge transfer is considered. We find $\Delta$=6.0 eV to lead to best agreement with the experimental data.  We extract charge transfer configurations of 89.7\% 3d$^{6}$ plus 10.3\% 3d$^{7}$\underline{L} for Fe$^{2+}_{oct}$, 90.4\% 3d$^{5}$ plus 9.6\% 3d$^{6}$\underline{L} for Fe$^{3+}_{oct}$, and an 89.6\% 3d$^{5}$ plus 10.4\% 3d$^{6}$\underline{L} for Fe$^{3+}_{tet}$. At the Fe L$_3$-edge the calculated XMCD exhibits the typical three peaks  expected for the inverse spinel structure of magnetite indicating the antiferromagnetic alignment of the octahedrally coordinated Fe$^{2+}_{oct}$ (feature A) and Fe$^{3+}_{oct}$ (C) ions with the tetrahedrally coordinated Fe$^{3+}_{tet}$ ions (B). The sum of the three calculated spectra, which may be shifted as to the energy scale since chemical bonding and band-like effects are not considered by the localized model,\cite{kui97} is used to fit the experimental spectra and estimate the cationic distribution of each magnetite thin film.

\begin{figure}[hbt]
    \centering
        \includegraphics[width=8.125cm]{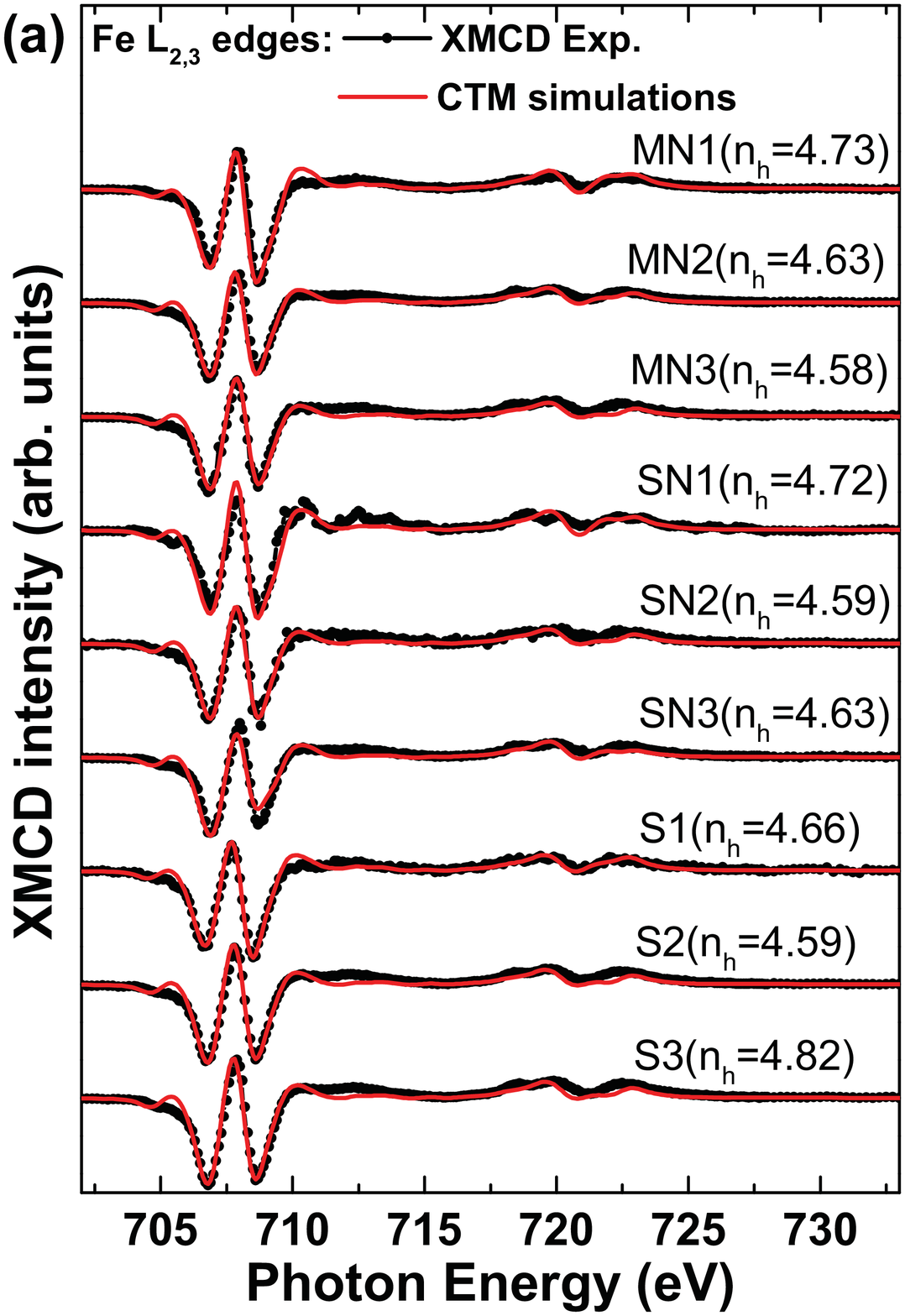}
        \includegraphics[width=8.125cm]{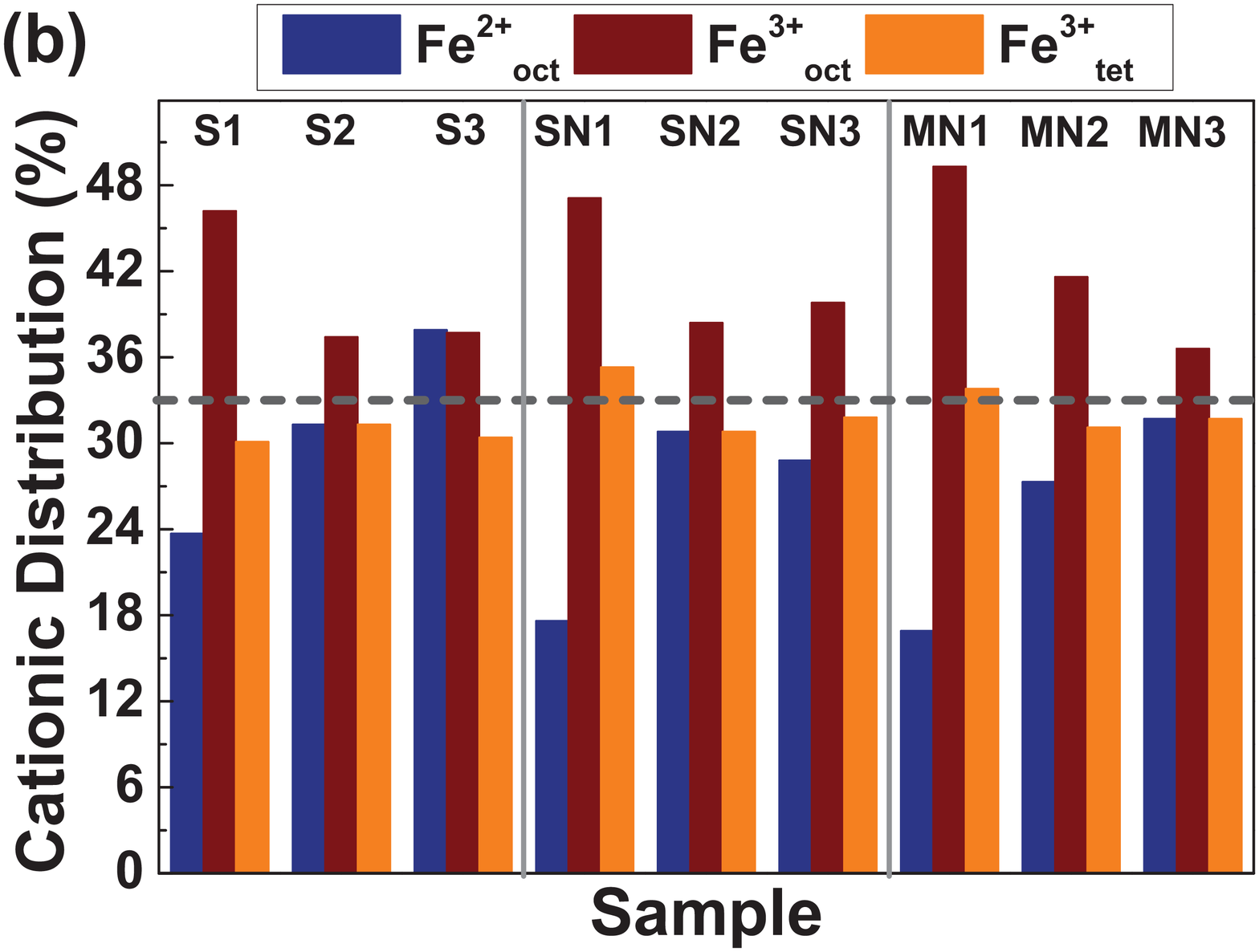}
    \caption{\label{Fig6} (a) Fe L$_{2,3}$-XMCD spectra and the corresponding fits from CTM simulations for all samples investigated in this work. The average number of holes (n$_h$) per Fe atom as derived from the CTM simulations is also given in parentheses. (b) Cationic distribution of the divalent and trivalent iron ions in octahedral and tetrahedral coordination as derived from the CTM simulations.}
\end{figure}

Fig.~\ref{Fig6}a displays the Fe L$_{2,3}$ XMCD spectra and the corresponding fits for all samples discussed in this work. The cation distributions found are presented in a diagram (Fig.~\ref{Fig6}b). We see that the iron cation distribution is significantly different from ideal magnetite for samples \textbf{S1}, \textbf{SN1}, \textbf{MN1}, i.e.\ those samples with the thinnest iron oxide layers. These samples exhibit a lack of Fe$^{2+}_{oct}$ but an excess of Fe$^{3+}_{oct}$ at the surface. With increasing film thickness of the iron oxide layers the cation distribution converge towards that of stoichiometric Fe$_3$O$_4$, whereas a slight excess of Fe$^{3+}_{oct}$ ions at the surface near region remains.
The excess of Fe$^{3+}_{oct}$ found in particular in the thinner films can likely be attributed to maghemite ($\gamma$-Fe$_2$O$_3$) formation at the magnetite surface. Such an effect has been very recently investigated by Fleischer \emph{et al.}\cite{fle14,sch15b} employing Raman spectroscopy. Fleischer \emph{et al.}\ found on uncapped 4 nm Fe$_3$O$_4$ films on MgO(001) that most of the maghemite formation occurs within days under ambient conditions and then slows significantly down in the time-frame of months.\cite{fle14} As to the samples studied in this work, series \textbf{S} has been grown around 4 month, series \textbf{SN} and \textbf{MN} around 2-3 weeks prior to the XMCD experiments. The HAXPES experiments have been performed quite some time (over one year) after the thin film synthesis. Our experiments confirm a potential maghemite formation on the magnetite surfaces as to the cationic distribution determined from the XMCD experiments and the HAXPES spectra recorded at lower excitation energies. This applies in particular to the thin magnetite layers ($\sim$5-10 nm). However, thicker Fe$_3$O$_4$ thin films appear to be much more stable also at their surfaces. This is obviously also the fact a couple of month and over one year after synthesis (see e.g.\ Fe L$_{2,3}$-XMCD analysis of samples \textbf{S2} and \textbf{S3} and HAXPES results of sample \textbf{S2}). Furthermore the total electron yield used to measure the XMCD spectra is known to be rather surface sensitive in the soft x-ray regime. For the Fe L$_{2,3}$ resonances of iron oxides the probing depth has been estimated to be around 0.8-3 nm.\cite{goe06,gom14} As the thinnest magnetite layers investigated here are around 6 nm (\textbf{MN1} and \textbf{SN1}) we can assume that the XA- and XMCD spectra stem from the first few nm and are not influenced by the substrate or the interface to the NiO buffer layer. Hence, we find a very different rate of maghemite formation within the first few 2-3 nm of the magnetite layers in dependence of the overall film thickness. Within the limits of the experiment, Fe$_3$O$_4$ layers with thickness $\geq$ 20 nm exhibit an almost stoichiometric magnetite cation distribution, independently from the substrate or the presence of a NiO buffer layer (see also Fig.~\ref{Fig6}b). Similar results have been recently reported for magnetite thin films directly deposited in MgO(001) substrates.\cite{sch15b}

Next we will discuss the magnetic spin moments determined by the spin sum rule. We limit our analysis to the spin magnetic moment which represents approximately 90\% or more of the overall magnetic moment in iron oxides. We want to note that the orbital moment in magnetite is still under discussion in the literature,\cite{hua04,goe06,mar15} but a precise determination would require more intricate experiments and an extended energy range, which are beyond the scope of this work.

\begin{figure}[hbt]
    \centering
        \includegraphics[width=8.125cm]{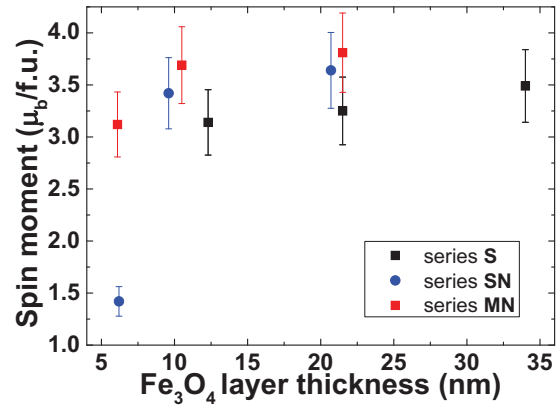}
    \caption{\label{Fig7} Calculated magnetic spin moments in $\mu_B$/f.u. from the XMCD spin sum rule.}
\end{figure}

In order to extract the spin magnetic moments we use the spin sum rule developed by Chen \emph{et al.}\cite{che95} The number of holes are determined from the charge transfer multiplet simulations for each sample. We also account for the core hole interactions which mix the character of the L$_3$ and L$_2$ edges\cite{ter96,pia09} by considering the spin sum rule correction factors obtained by Teramura \emph{et al.}\cite{ter96} The spin magnetic moments are depicted in Fig.~\ref{Fig7}. One can see that an increasing magnetite layer thickness results in a higher spin moment. Samples \textbf{S2}, \textbf{SN3} and \textbf{MN3} exhibit spin magnetic moments of 3.3, 3.6, and 3.8 $\mu_{B}$/f.u., respectively. These results (excluding the orbital magnetic moment) are rather close to the bulk magnetic moment of Fe$_3$O$_4$.\cite{wei29} The thinner films have decreased spin moments, which might be partly due to maghemite formation on the very surface layer of these films. However, this cannot explain in particular the rather low spin moment found for sample \textbf{SN1} (which exhibits a very similar cationic distribution as sample \textbf{MN1}), 1.45 $\mu_{B}$/f.u., which is the same value reported for a 2.5 nm Fe$_3$O$_4$ thin film on BaTiO$_3$.\cite{bab13} One possible reason for this might stem from epitaxial strain.\cite{kus15} However, as well as sample \textbf{SN1} also samples \textbf{SN2} and \textbf{SN3} show slight tensile strain\cite{kus15} of less than 1\% compared to the bulk value of Fe$_3$O$_4$ and have spin magnetic moments close to that found for the respective samples from series \textbf{MN}, the latter ones exhibit slight compressive strain\cite{kus15} of less than 1\% in vertical direction. Another potential reason for this reduced spin moment might be connected with the higher roughness of the Fe$_3$O$_4$ layer which is found to be 0.45 nm for sample \textbf{SN1}, since all other magnetite layers from series \textbf{SN} and \textbf{MN} have lower roughnesses of approximately 0.3 nm.\cite{kus15} The spin moment of sample \textbf{MN1} (3.1 $\mu_{B}$/f.u.) is comparable to that very recently reported for Fe$_3$O$_4$(001) thin films grown on Ag(001) of similar thicknesses.\cite{gom14} Except for sample \textbf{SN1} we find significant higher spin magnetic moments for our thin films than observed for similar (001) oriented Fe$_3$O$_4$ films grown on MgO(001) substrates.\cite{orn10}

\section{Summary}
In summary, we performed a comprehensive electronic and magnetic structure investigation of thin magnetite films and NiO/Fe$_3$O$_4$(001) bilayers grown on MgO(001) and Nb-SrTiO$_3$(001) by means of RMBE. In-situ characterization using LEED and XPS shows formation of epitaxial and stoichiometric NiO layers and Fe$_3$O$_4$ thin films. Thickness of the NiO layers is varied between 6 and 12nm and that of the Fe$_3$O$_4$ layers between 5 and 33 nm as to XRR measurements. Further electronic and chemical properties are tackled by HAXPES experiments with variable excitation energy. All thin films appear to comprise stoichiometric magnetite and NiO in deeper layers and at the interface. However, thinner Fe$_3$O$_4$ films (below 10 nm) tend to form Fe$^{3+}$ on the very surface, likely in form of maghemite. MOKE magnetization curves show significant increased magnetic coercive field and that the in-plane magnetic axis is rotated by 45$^{\circ}$. We employ element specific XMCD at the Fe L$_{2,3}$  edges to tackle the iron cationic distribution in detail. In agreement with the HAXPES we find a lack of Fe$^{2+}$ at the surface of in particular the thinner magnetite layers investigated in the framework of this study. Thicker films are close to stoichiometric cation distribution also at their surface and exhibit spin magnetic moments close to that known from bulk material, whereas thinner samples ($<$ 10nm) show only somewhat reduced spin moments. The only exception is the sample with 5 nm NiO and 6 nm Fe$_3$O$_4$ bilayer grown on Nb-SrTiO$_3$(001) (\textbf{SN1}), here the strongly reduced spin moment might be due to an increased roughness compared to the magnetite layers of the other samples investigated here. Angular dependent MOKE experiments reveal the magnetic easy axis to be along the $\langle$100$\rangle$ thin film directions.\par
In conclusion we demonstrate that the interface of ferrimagnetic Fe$_3$O$_4$ and antiferromagnetic NiO leads to a low concentration of Fe $t_{2g}$ spin down states present directly at the interface. The $\gamma$-Fe$_{2}$O$_3$ formation seems to be less intense on the surface of thicker (uncapped) magnetite layers ($\geq$ 20nm) than found for thin layers even in time frames of several months of one year under ambient conditions which might be interesting for potential applications. The altered electronic and magnetic properties as the density of states near Fermi level, magnetic coercive field and anisotropy deserve further future investigations. An improved understanding could open an avenue of tailoring the desired magnetic properties by choosing an appropriate set of substrate, layer thicknesses, and preparation parameters.

\begin{acknowledgments}
Financial support by the Deutsche Forschungsgemeinschaft (DFG) (KU2321/2-1) is gratefully acknowledged.
Part of this work has been performed at the Advanced Light Source, ALS, Lawrence Berkeley National Laboratory, Berkeley, USA; which is operated under contract No. DE-AC03-76SF00098. We acknowledge the Helmholtz-Zentrum Berlin for provision
of synchrotron radiation beam time at beamline KMC-1 of Bessy II.
\end{acknowledgments}

\bibliography{Fe3O4-NiO_MCD}% Produces the bibliography via BibTeX.

\end{document}